\documentstyle[12pt,epsf,epsfig]{article}
\newcommand{\bp}{{\bf p}}
\newcommand{\bk}{{\bf k}}
\newcommand{\bx}{{\bf x}}
\newcommand{\by}{{\bf y}}
\newcommand{\bz}{{\bf z}}
\newcommand{\bzz}{{\bf 0}}
\newcommand{\Tr}{{\mbox{Tr}}}
\newcommand{\btab}{\begin{tabbing}}
\newcommand{\etab}{\end{tabbing}}
\newcommand{\eqntimes}{\mbox{} \times}

\newcommand{\beqn}{\begin{equation}}
\newcommand{\eeqn}{\end{equation}}
\newcommand{\barr}[1]{\begin{array}{#1}}
\newcommand{\earr}{\end{array}}
\newcommand{\beqna}{\begin{eqnarray}}
\newcommand{\eeqna}{\end{eqnarray}}
\newcommand{\btablec}{\begin{table} \begin{center}}
\newcommand{\etablec}{\end{center} \end{table}}

\newcommand{\gapproxeq}{\lower.7ex\hbox{$\;\stackrel{\textstyle>}
{\sim}\;$}}
\newcommand{\plabel}[1]{\label{#1}}
\newcommand{\pbibitem}[1]{\bibitem{#1}}
\input epsf

\begin{document}
\title{
\begin{flushright} 
\end{flushright} 
\vspace{0.6cm}  
\Large\bf (Field) Symmetrization Selection Rules}
\vskip 0.2 in
\author{Philip R. Page\thanks{\small \em E-mail:
prp@lanl.gov}
\\{\small \em  Theoretical Division, Los Alamos
National Laboratory, Los Alamos, NM 87545, USA}}
\date{}
\maketitle
\begin{abstract}{QCD and QED exhibit an infinite set of three--point Green's 
functions that contain only OZI rule violating contributions, 
and (for QCD) are subleading in the large $N_c$ expansion. 
The Green's functions describe the ``decay'' of a 
$J^{PC}=\{1,3,5\ldots\}^{-+}$ exotic hybrid meson current to two 
$J=0$ (hybrid) meson currents with identical $P$ and $C$.
We prove that the QCD amplitude for a neutral hybrid
$\{1,3,5\ldots\}^{-+}$ exotic current to create 
$\eta\pi^{0}$ only comes from OZI rule violating contributions under 
certain conditions, and 
is subleading in $N_c$.
}
\end{abstract}
\bigskip


PACS number(s): 11.15.Pg \hspace{.2cm}11.15.Tk \hspace{.2cm} 
11.30.Na \hspace{.2cm}11.55.-m
\hspace{.2cm} 12.38.Aw \hspace{.2cm} 12.39.Mk \hspace{.2cm} 13.25.Jx

\vspace{.3cm}


\section{Introduction}

More than a decade ago Lipkin argued that an explicitly identified 
contribution to the 
decay of a $J^{PC}=\{1,3,5\ldots\}^{-+}$ ``exotic'' hybrid meson to $\eta\pi$
vanishes~\cite{lipkin}. 
Here $J$ denotes the internal angular momentum,
$P$ (parity) the reflection through the origin and 
$C$ (charge conjugation) particle--antiparticle exchange. These
are conserved quantum numbers of the strong and electromagnetic interactions.
By ``hybrid meson'' we mean a fermion--antifermion with additional gauge 
bosons. 
``Exotic'' means that the $J^{PC}$ cannot be constructed
for conventional mesons in the quark model, or equivalently that there
are no local currents built only from a fermion and antifermion field that 
can have these $J^{PC}$.

Lipkin's intuitive argument was later extended and
placed in a more formal context, called
``symmetrization selection rules''~\cite{sel}.
However, these need to be converted into rigorous
quantum field theoretic arguments, which is the subject of the current work.

We outline Lipkin's intuitive argument for the decay of a positively charged
$J^{P} = 1^{-},\; 3^{-},\; \ldots$ state to $\eta\pi^+$ when
strong interactions with
isospin symmetry is assumed. G-parity conservation implies that
the neutral isospin partner of the initial state is $J^{PC}$ exotic.
We consider the decay process where a quark and antiquark in the initial
state proceed in such a way that the quark ends up in the one final meson,
and the antiquark in the other meson (Fig. \ref{lip}). This decay process
is called ``connected'', because each meson is connected to the other mesons
via quark lines. The gluons are not indicated. The fact that the
neutral isospin partner of the initial state is $J^{PC}$ exotic, and
that the initial state contains a quark and antiquark,
implies that the initial state is a hybrid meson.
The argument is depicted in Fig. \ref{lip}. Taking the initial hybrid
at rest, the $\eta$ and $\pi^+$ emerge with momenta $-\bf k$ and 
$\bf k$ respectively. First consider the three top left diagrams. The top
diagram has a negative sign in front by convention.
When the transformation ${\bf k} \leftrightarrow -{\bf k}$ is applied,
the middle diagram is obtained, noting that the decay is in an odd partial
wave, which acquires a minus sign under the transformation. 
This is a general property of
odd partial waves. The bottom diagram is obtained by noting that
the amplitude to create a $u\bar{u}$ pair is the same as for a
$d\bar{d}$ pair by the assumption of isospin symmetry, which treats the
up and down quarks the same. 
The three top right ``hadronic'' diagrams are now obtained from the
three top left ``quark'' diagrams by attaching the initial hybrid to the
initial $u\bar{d}$ quarks, and the final $\pi^+$ to the final 
$u\bar{d}$ quarks. Since the flavour wave function of the $\eta$ is
proportional to $u\bar{u}+d\bar{d}$, it is attached to either $u\bar{u}$
or $d\bar{d}$, with a positive relative sign. Because each of the
three top left quark  
diagrams are equal, it follows that each of the three
top right hadronic  
diagrams are equal. The two bottom diagrams depict the decay
amplitude, taking into account that there are two possible ways for the
final $\eta$ and $\pi^+$ to couple, since the quark in the initial state
can go either to the $\eta$ or the $\pi$. Looking back at the
top right hadronic diagrams one immediately notices that the decay 
amplitude vanishes. This is the desired result.

\begin{figure}[t]
\begin{center}
\vspace{-1.3cm}
\hspace{-0.5cm}\epsfig{file=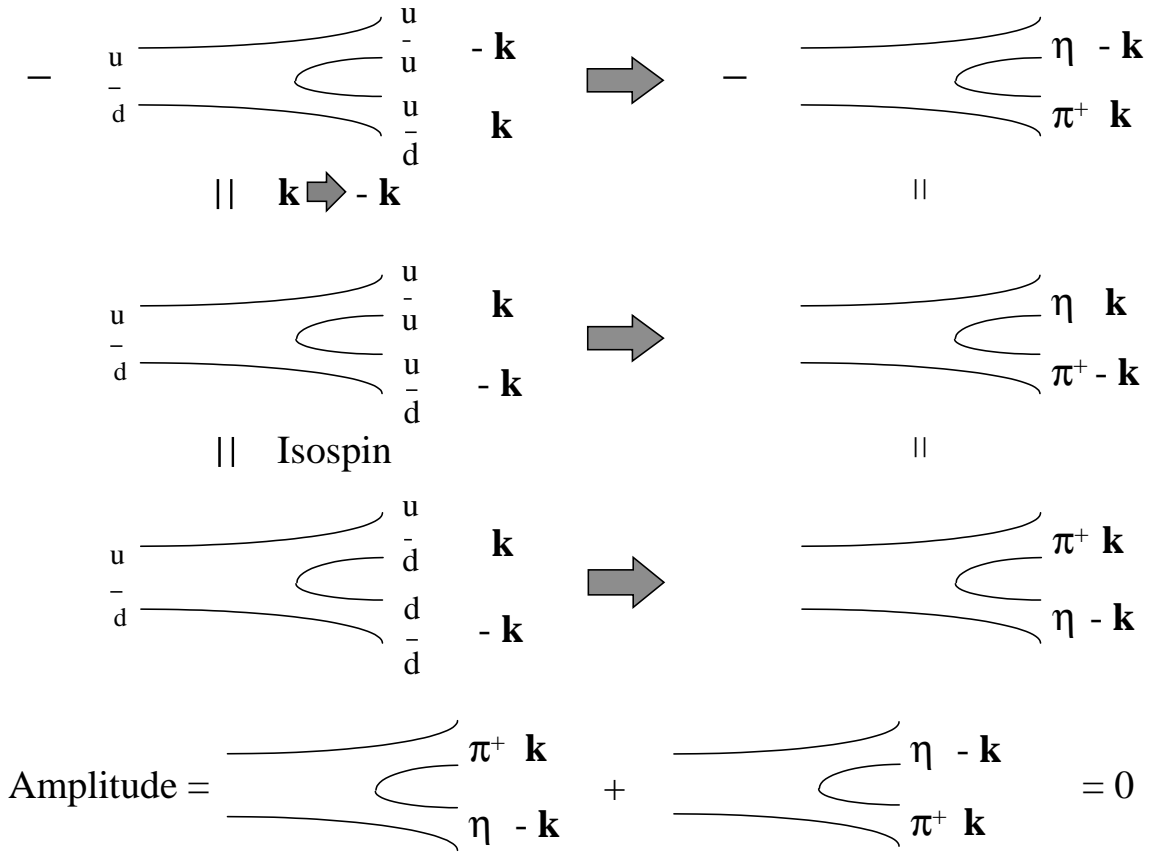,width=17cm,angle=0}
\vspace{-1cm}
\caption{\plabel{lip} Lipkin's intuitive argument.}
\end{center}
\end{figure}

The argument above can be repeated without assuming isospin symmetry,
as will be the case in the remainder of this work, if the initial
state is neutral. Hence isospin is not an essential assumption.
It appears strange that a decay amplitude vanishes from very 
general considerations if the decay is allowed by the conserved quantum
numbers of the strong interaction. This paradox is resolved when one
notices that it was only argued that the connected contribution to
the decay vanishes, not the entire decay amplitude. Lipkin's argument
serves as a guide to how a quantum field theoretic argument would proceed.

In this Paper we demonstrate that some explicitly identified contributions to
certain three--point Green's functions vanish.
This is called {\it field symmetrization
selection rules}. Particularly,
the connected (Okubo--Zweig--Iizuka (OZI) allowed) 
contribution to each Green's function vanishes. 
Hence, the Green's function only has a disconnected (OZI forbidden)
contribution, which is expected to be 
phenomenologically suppressed by virtue of the
OZI rule~\cite{ozi}. 
The Green's function is built from two 
$J=0$ (hybrid) 
meson currents with identical $P$ and $C$, and a 
$J^{PC}=\{1,3,5\ldots\}^{-+}$ exotic hybrid
meson current. 

We subsequently investigate the physical consequences. The amplitude
for a $\{1,3,5\ldots\}^{-+}$ hybrid current to create 
$\eta\pi^0$ is shown to be proportional to the Green's function 
under certain conditions.
Because some explicitly identified contributions to
 the Green's function vanish, it follows
that the amplitude does not come from these contributions, 
called {\it symmetrization selection
rules at the hadronic level}.
Particularly, the amplitude does not arise from the connected contribution to
the Green's function, and hence only from the disconnected contribution,
which is expected to be suppressed by virtue of the OZI rule.
The experimental consequences
pertain to a central issue in hadron spectroscopy: the search for
hybrid meson bound states beyond conventional mesons and baryons.

For Quantum Chromodynamics (QCD) with a large number of colours $N_c$,
the foregoing conclusions will be expressed as follows. Because a disconnected
contribution to a Green's function is subleading in the large $N_c$ expansion,
and the Green's function only has a disconnected contribution, the 
Green's function is subleading in $N_c$. Also, since the amplitude
for a $\{1,3,5\ldots\}^{-+}$ hybrid current to create $\eta\pi^0$ only 
comes from the disconnected contribution to the Green's function, it is
subleading in $N_c$.

It has previously been shown that the connected part 
of the {\it quenched Euclidean} three--point Green's function of 
specific hybrid meson neutral $1^{-+}$ and two pseudoscalar ($J=0$) currents
vanishes exactly in QCD, if {\it isospin} symmetry is assumed 
\cite{pene}. We remove the three italized superfluous assumptions, but still
need the connected part and hybrid meson currents. We also generalize 
beyond specific currents, beyond
$1^{-+}$ and pseudoscalar currents, beyond the connected part,
and beyond QCD.

In section 2 the currents and Green's functions are introduced. The 
principle of symmetrization is developed. In section 3 the Green's 
functions are calculated leading to succinctly stated 
field symmetrization selection rules. An explicit example is discussed.
In section 4 the physical consequences are investigated, yielding
symmetrization selection rules at the hadronic level, which are
concisely stated. Section 5 studies the selection rules in the 
large $N_c$ expansion. Section 6 contains further remarks.

\section{Symmetrization}

Consider local currents of the form

\beqn\plabel{cur}
\{A_{\mu}(x),B(x),C(x)\} = \sum_i \bar{\psi}_i(x)\;\{a_i a_{\mu}(x),b_i b(x),c_i c(x)\}\; {\psi}_i(x)
\eeqn
where ${\psi}_i(x)$ is a quark or lepton
 field of flavour $i$, and $a_i,b_i,c_i$ are c--numbers
weighting the flavours. The currents are diagonal in flavour.
The ``matrices'' $a_\mu(x),\; b(x)$ and $c(x)$
 contain an arbitrary number of Dirac matrices, 
Gell--Mann colour matrices, derivatives
(acting both to the left and the right), 
gluon or photon fields and correspond to gauge 
invariant currents~\cite{gen}. 
A common choice for the flavour structure of
the three currents is  $\bar{u}u-\bar{d}d,\; 
\bar{u}u+\bar{d}d - \bar{s}s$ and $\bar{u}u-\bar{d}d$,  
interpolating for an isovector resonance, an $\eta$ and a $\pi^0$.
The matrix $a_{\mu}(x)$ in Eq.~\ref{cur} is chosen to ensure
that $P=-$. We require that currents $B$ and $C$ have $J=0$, 
as well as equality of $b(x)$ and $c(x)$, implying equal $P$ and $C$. 
Note that equality of the matrices $b(x)$ and $c(x)$ does {\it not} imply
equality of the currents $B(x)$ and $C(x)$.
Because $C$ is the same for both these currents, charge conjugation
conservation requires that the $C$ of $A_{\mu}(z)$ is $+$. 
Take  $A_\mu$ to have odd $J$~\cite{odd}.
Hence $A_{\mu}$ must be chosen to have $J^{PC}=\{1,3,5\ldots\}^{-+}$.
These are exotic quantum numbers, and $A_{\mu}$ is built from a 
fermion and an antifermion field, so that the current is 
a hybrid current, i.e. has to contain at least one gluon or photon field.
For the current with $J\neq 0$, the 
appropriate Lorentz indices are indicated by $\mu$.

We start by demonstrating that certain three--point
Green's functions are equal to their antisymmetric parts.
This is done by first defining
the spatial Green's function, $G_{\mu}(x,y,z)$, and 
decomposing it into symmetric and 
antisymmetric parts. The Green's function of interest will be the
Fourier transform of the spatial Green's function, $G_\mu(\bp,t)$. 
We then argue that
the Fourier transform of the symmetric part of the spatial Green's function,
$G_\mu^S(\bp,t)$, vanishes. Hence, $G_\mu(\bp,t)$ equals the
Fourier transform of the antisymmetric part of the spatial Green's function,
$G_\mu^A(\bp,t)$.

Define the three--point Minkowski space Green's function

\beqn\plabel{sym}
G_{\mu}(x,y,z) =\langle  0 |\; B(x)\;C(y) \;A_{\mu}(z)\; |0\rangle 
\eeqn
with $x_0 = y_0\equiv t, \; z_0=0$ and $\bz=\bzz$. 
The Green's function describes the decay (production) process of a
current $A_{\mu}(z)$ at time 0 propagating into the final (initial) currents
$B(x)$ and $C(y)$ at some positive (negative) time $t$.
Although we shall refer to $G_{\mu}(x,y,z)$ as a ``Green's function'', the 
usual usage of the term requires the currents to be at different times, 
i.e. $x_0 \neq y_0$, with the currents ordered from positive to
negative times.
The Green's function $G_{\mu}(x,y,z)$
can be written as $G_{\mu}^S(x,y,z)+G_{\mu}^A(x,y,z)$,
i.e. the sum of parts symmetric and antisymmetric under exchange of
$x$ and $y$, as any function of $x$ and $y$ can be written.

Define the Fourier transforms

\beqna\plabel{def}
{ G_\mu}(\bp,t) &\equiv &\int\; d^3x\; d^3y\; e^{i(\bp\cdot\bx-\bp\cdot\by)}\;
{ G_\mu}(x,y,z)   \\
\left\{ G_\mu^S(\bp,t),\; G_\mu^A(\bp,t) \right\}
&\equiv &\int\; d^3x\; d^3y\; e^{i(\bp\cdot\bx-\bp\cdot\by)}\;
\left\{ G_\mu^S(x,y,z), \; G_\mu^A(x,y,z)\right\} \plabel{adef}  
\eeqna
Note that these Fourier transforms
 cannot be inverted to give the spatial Green's functions,
since they only have one momentum variable $\bp$.
From Eqs. \ref{def} and 
\ref{adef} it follows that $G_{\mu}(\bp,t)= G_\mu^S(\bp,t) +
G_\mu^A(\bp,t)$.

By exchanging integration variables $\bx \leftrightarrow \by$

\beqna\plabel{f1}\lefteqn{\hspace{-11.1cm}
G_\mu(-\bp,t) = \int\; d^3x\; d^3y\; e^{i(\bp\cdot\bx-\bp\cdot\by)}\;
{ G_\mu}((\by,t),(\bx,t),z) \nonumber } \\
= \int\; d^3x\; d^3y\; e^{i(\bp\cdot\bx-\bp\cdot\by)}\; \left\{
G_\mu^S(y,x,z) + G_\mu^A(y,x,z) \right\} =
G_\mu^S(\bp,t) - G_\mu^A(\bp,t)  
\eeqna

Exchanging integration variables $\bx \rightarrow -\bx$ and
$\by \rightarrow -\by$ yields

\beqn\plabel{f2}
G_\mu(-\bp,t) = -G_\mu(\bp,t)
\eeqn
by using that ${G_\mu}((-\bx,t),(-\by,t),z) 
= -{G_\mu}(x,y,z)$, by conservation of parity. We henceforth restrict
ourselves to the parity conserving theories of Quantum Electrodynamics
(QED) and QCD. We also used that the product of the parities of the
currents $B(x)$, $C(y)$ and $A_\mu(z)$ is $-1$, which follows from the
assumptions below Eq. \ref{cur}. This means that the decay
(production) process is in odd partial wave. Eq. \ref{f2} is a 
well--known property of such a process.

Combining Eqs. \ref{f1} and \ref{f2} yields the result

\beqn\plabel{fssr}
G^S_\mu(\bp,t) = 0 
\eeqn
Bose symmetry states that identical bosons are not allowed in an odd partial 
wave. The way one shows this would be analogous to the steps above
if $G^A_\mu(\bp,t) = 0$.
Hence, the Fourier transform of the symmetric part of 
$G_{\mu}(x,y,z)$ in Eq. \ref{fssr} vanishes by arguments that are the field
theoretical version of Bose symmetry. 

From Eq. \ref{fssr} follows the desired result

\beqn\plabel{finr}
G_\mu(\bp,t) = G^A_\mu(\bp,t) 
\eeqn

\section{Field Symmetrization Selection Rules}

We proceed to explicitly identify some contributions to  
$G_\mu(\bp,t)$ that vanish. 
This is attained via 
partially evaluating $G_\mu^A(\bp,t)$. It is subsequently shown
that the action of a certain operator, $\hat{O}_{\bp}$, on the
Fourier transform of the antisymmetric part of the Green's function,
$\hat{O}_{\bp}\; G_{\mu}^A(\bp,t)$, does not contain some contributions.
Because $ G_{\mu}^A(\bp,t) = G_{\mu}(\bp,t)$ from Eq. \ref{finr}, it follows
that $\hat{O}_{\bp}\; G_{\mu}(\bp,t)$ does not contain these contributions.

We now evaluate the contributions to the Green's function $G_\mu^A(x,y,z)$ 
with the currents in Eq.~\ref{cur}, using that
$G_\mu^A(x,y,z) = \frac{1}{2} \; (G_\mu(x,y,z) - G_\mu(y,x,z))$,

\beqna G_\mu^A(x,y,z) &=& \frac{1}{2} \;
\langle  0 |\; \left(\; B(x)\;C(y) - B(y)\;C(x) \; \right) \;A_{\mu}(z)\; 
|0\rangle
\nonumber  \\ 
& = & \frac{1}{2} \; \sum_{i\; j} b_i c_j \; 
\langle  0 |\; \left(\; \bar{\psi}_i(x)\; b(x)\; {\psi}_i(x) \;\;
\bar{\psi}_j(y)\; c(y)\; {\psi}_j(y)\;  - \right. \nonumber \\
&&\hspace{2.8cm}\left. \bar{\psi}_i(y)\; b(y)\; {\psi}_i(y) \;\;
\bar{\psi}_j(x)\; c(x)\; {\psi}_j(x) \; \right) \;A_{\mu}(z)\; |0\rangle
\plabel{dotti}
\eeqna
The various contributions to this expression are now
discussed.
Consider contributions to the expression where the same flavours
are isolated in the currents
$B(x)$ and $C(y)$, i.e. contributions where $i=j$. Using Eqs. \ref{adef}
and \ref{dotti},
these contributions to $G_\mu^A(\bp,t)$ can be written

\beqna\plabel{uit}\lefteqn{\hspace{-10cm}
G_\mu^A(\bp,t) \sim 
\int\; d^3x\; d^3y\; e^{i(\bp\cdot\bx-\bp\cdot\by)}\;
\frac{1}{2} \; \sum_{i} b_i c_i \; 
\nonumber } \\ \eqntimes
\langle  0 |\; \left[\; \bar{\psi}_i(x)\; b(x)\; {\psi}_i(x)\; , \;
\bar{\psi}_i(y)\; b(y)\; {\psi}_i(y) \; \right]\; A_{\mu}(z)\;|0\rangle
\equiv \Lambda_\mu(\bp,t) 
\eeqna
where we used $b(x) = c(x)$, and denoted the contributions 
by $\Lambda_\mu$. The important observation is that the difference
of currents in Eq. \ref{dotti} has simplified to the commutator of
currents in Eq. \ref{uit}. 
It is possible to show that $\Lambda_\mu$ has a polynomial 
dependence on $\bp$ (see Appendix)~\cite{dir}. 
There hence exists a polynomial
operator (containing a derivative of high enough power in $\bp$)
with the property that
$\hat{O}_{\bp}\; \Lambda_\mu(\bp,t) = 0$.
This result will later be demonstated in an explicit example.
We conclude from Eq. \ref{uit} 
that $\hat{O}_{\bp} \; G_\mu^A(\bp,t)$, and from Eq. \ref{finr} also
$\hat{O}_{\bp} \; G_\mu(\bp,t)$, do not contain
contributions from the same flavour in 
currents $B(x)$ and $C(y)$, a result which corrects the former
treatment~\cite{ft}.
Although this is the desired result, and the end of the mathematical 
derivation, it will be pivotal to 
develop a more intuitive understanding of the contributions.

$\hat{O}_{\bp} \; G_\mu(\bp,t)$ is an operator acting on the Fourier transform
of $G_\mu(x,y,z)$. 
For the purpose of illustration in the next few paragraphs, 
consider the time $t$
to be positive, with 
$B(x)$ slightly advanced at time $x_0 = t+\delta t$
with respect to $C(y)$ at time $y_0=t$  
in the definition of $G_\mu(x,y,z)$ in Eq. \ref{sym}.
Here $\delta t$ is small, positive and non-zero. 
The sign of $t$ or the magnitude of $\delta t$ will not change our 
eventual conclusions.
Since all the 
currents are at different times, with the currents ordered from large to
small times, it follows that $G_\mu(x,y,z)$ is a Green's function
according to the usual usage of the term, which implies that it can be
represented by a path integral. For concreteness, consider the
contribution to $G_\mu(x,y,z)$ in Eq.~\ref{sym} from the up quark flavours in
the currents $B(x)$, $C(y)$ and $A_\mu(z)$ in Eq.~\ref{cur}, which is
(modulo $a_u b_u c_u$) 

\beqna\plabel{path}\lefteqn{
\int{\cal D}A\;{\cal D}\bar{\psi}\;{\cal D}\psi\;\delta(f(A))\; \mbox{det}{\cal M}_F\;  \bar{u}(x)b(x)u(x)\;  \bar{u}(y)c(y)u(y)\;  \bar{u}(z)a_\mu(z)u(z) \; \exp (-i\int d^4x \; {\cal L}) \nonumber } \\ & & \hspace{-.6cm}=
\int\;{\cal D}A\;\delta(f(A))\; \mbox{det}{\cal M}_F\;  \mbox{det}\gamma_0 S^{-1}\;  \exp (-i\int d^4x \; {\cal L}^{(A)})\;
\nonumber \\ & & \hspace{-.8cm}
\{ \Tr[ 
-a_\mu(z)\; S_u^{A}(z,x)\; b(x)\; S_u^{A}(x,y)\; c(y)\; S_u^{A}(y,z) 
-a_\mu(z)\; S_u^{A}(z,y)\; c(y)\; S_u^{A}(y,x) \; b(x)\; S_u^{A}(x,z) ]
\nonumber \\ & & \hspace{-.6cm}
+\; \Tr [a_\mu(z)\; S_u^{A}(z,x)\; b(x)\; S_u^{A}(x,z)]\; \Tr[c(y)\; S_u^{A}(y,y)]
\; +\; \Tr [a_\mu(z)\; S_u^{A}(z,y)\; c(y)\; S_u^{A}(y,z)]\; 
\nonumber \\ & & \hspace{-.6cm}\eqntimes\;\Tr[b(x)\; S_u^{A}(x,x)]
\; +\; \Tr [a_\mu(z)\; S_u^{A}(z,z)] \; \Tr[b(x)\; S_u^{A}(x,y)\; c(y)\; S_u^{A}(y,x)]
\nonumber \\ & & \hspace{-.6cm}
-\;\Tr[a_\mu(z)\; S_u^{A}(z,z)]\; \Tr[b(x)\; S_u^{A}(x,x)]\; \Tr[c(y)\; S_u^{A}(y,y)]
\}
\eeqna
The gauge--fixing condition
$\delta(f(A))$ and Faddeev--Popov determinant $\mbox{det}{\cal M}_F$
are indicated. The QCD / QED Lagrangian is denoted by 
${\cal L}$ and the part containing terms without fermions ${\cal L}^{(A)}$.
The up quark propagator in a background field is $S_u^{A}(x,y)$ and
the fermion determinant (containing fermion 
loops) $\mbox{det}\gamma_0 S^{-1}$~\cite{prp}.

\begin{figure}
\begin{center}
\leavevmode
\hspace{-.3cm}\hbox{\epsfxsize=5 in}
\epsfbox{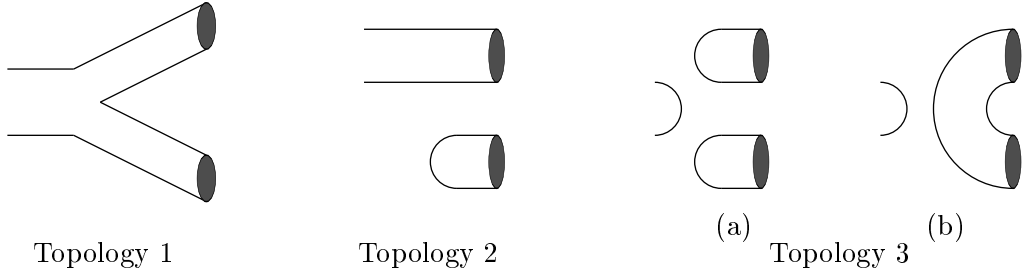}
\vspace{1.2cm}\caption{\plabel{topo}
Topologies contributing to Eq. \protect\ref{path}.}\end{center}
\end{figure}

The terms on the right hand side (R.H.S.)
of Eq. \ref{path} correspond to the topologies in Fig. \ref{topo}.
This is seen by associating with each up quark propagator 
an up quark line in the figure. Also, $z$ is associated
with the left hand side (L.H.S.) of each topology, and $x,\; y$ with the
blobs on the R.H.S.
In Eq. \ref{path} the first two terms correspond to
the quark line ``connected'' topology 1 in Fig. \ref{topo}, 
the second two to 
topology 2, term
five to topology 3b and term six to topology 3a.

The topologies hence represent all the different ways that a fermion and
antifermion field on the L.H.S. of Eq. \ref{path} can be ``contracted'' by
the fermion integration to yield the fermion propagators on the R.H.S.

Similar manipulations to Eq. \ref{path} can be performed for contributions
where not all the fermions in the currents
are up quarks, or not all the fermions have the
same flavour. 

From Fig. \ref{topo} it can be seen that 
topologies 1 and 3b have the interesting property
that when a certain flavour contributes from current $B(x)$,
then the {\it same} flavour contributes from current $C(y)$. 
This is because the currents are diagonal in flavour.
Although $B(x)$ and $C(y)$ contain various different flavour structures
according to Eq. \ref{cur}, topologies 1 and 3b force only the same
fermion flavours in $B(x)$ and $C(y)$ to contract.
For the
other topologies, it is sometimes the case that the flavours from 
two currents are the same, but not always. 
Recalling that $\hat{O}_{\bp} \; G_\mu(\bp,t)$ does not contain
contributions from the same flavour in 
currents $B(x)$ and $C(y)$, we derive

{\bf Field symmetrization selection rules (FSSR):} {\it
All contributions to the three--point Green's function 
$\hat{O}_{\bp}\; G_\mu(\bp,t)$
from the connected topology 1 and topology 3b (and some from topologies 2 
and 3a), i.e. contributions from the same flavour in 
currents $B(x)$ and $C(y)$,
vanish exactly for all momenta $\bp$ and times $t$
for an infinite
set of equal matrices $b(x) = c(x)$ and different flavour structure
currents $B(x)$ and $C(y)$ in QCD and QED.
The Green's function decribes the ``decay'' of a 
$J^{PC}=\{1,3,5,\ldots\}^{-+}$ exotic hybrid meson current to two
$J=0$ (hybrid) meson currents $B(x)$ and $C(y)$, which are identical
$(b(x)=c(x))$, e.g. have identical parity and charge conjugation,
except possibly for their flavour.}

The contributions that vanish by the FSSR are those from the same 
flavour in currents $B(x)$ and $C(y)$. Since the product $B(x)\; C(y)$
occurs in the Green's function (see Eq. \ref{sym}), these are contributions
of the form $b_ic_i\;\bar{\psi}_i(x)\; b(x)\; {\psi}_i(x) \;
\bar{\psi}_i(y)\; c(y)\; {\psi}_i(y)$, where $i$ indicates the flavour,
and no summation is implied. It is evident that the precise values of
$b_i$ and $c_i$, i.e. the flavour structure of $B(x)$ and $C(y)$, do not 
affect the fact that the contributions vanish by the FSSR. The same is
true for $A_\mu(z)$. For example, both the common choice of
$\bar{u}u-\bar{d}d,\; 
\bar{u}u+\bar{d}d - \bar{s}s$ and $\bar{u}u-\bar{d}d$ for the flavour
structure of $A_\mu(z)$, $B(x)$ and $C(y)$, and the alternative choice
of respectively $\bar{u}u-2\bar{d}d,\; 
\bar{u}u+\bar{d}d - 2\bar{s}s$ and $2\bar{u}u-\bar{d}d$, obey the
FSSR.

The FSSR do not depend on the various parameters of QCD and QED, 
i.e. masses, couplings, charges, number of colours and flavours, but do
require the $CP$ violating $\theta$ parameter to be zero, since this
parameter violates parity conservation. FSSR also occur
is pure QED, where QCD interactions are turned off.

We now discuss an example to which the FSSR apply.
The gauge invariant isovector--like local
current $A_{\mu}(z) = (T^i_{jk})_\mu\;\left (\bar{u}(z)F^{a}_{jk}(z)\lambda^a
\gamma_{i}u(z)-\bar{d}(z)F^{a}_{jk}(z)\lambda^a
\gamma_{i} d(z) \right )$
is $J^{PC}=1^{-+}$ exotic \cite{pene}, 
where $\gamma_{i}$ and $\lambda^a$ are
Dirac and Gell--Mann colour matrices respectively, and
$(T^i_{jk})_\mu$ is a tensor combining the spatial 
indices $i,j,k$ to build a spin 1 object.
It is a hybrid current,
since it contains a gluon field, as can be seen by the presence of
the gluon field tensor $F^{a}_{jk}$. The gauge--invariant
isoscalar-- and isovector--like local currents $B(x)=\bar{u}(x)\gamma_5 u(x)+
\bar{d}(x)\gamma_5 d(x)-\bar{s}(x)\gamma_5 s(x)$ and 
$C(y)=\bar{u}(y)\gamma_5 u(y)-\bar{d}(y)\gamma_5 d(y)$
are pseudoscalar ($J^{PC}=0^{-+}$), with $\gamma_5$ a Dirac matrix. Note that
$b(x)=c(x)=\gamma_5$, but that $B(x) \neq C(x)$. 
Also, we do not 
assume isospin symmetry. To evaluate $\Lambda_\mu$ in Eq. \ref{uit},
the commutator $[\bar{u}(x)\gamma_5u(x),\bar{u}(y)\gamma_5u(y)]$
must be evaluated. The commutator can be shown to equal 
$\delta^3(\bx-\by)\; \left(\: \bar{u}(x)\gamma_5\gamma_0\gamma_5u(y)-
\bar{u}(y)\gamma_5\gamma_0\gamma_5u(x)\:
\right)$. Because of the delta function, it is immediate that
$\Lambda_\mu(\bp,t)$ in Eq. \ref{uit} is independent of $\bp$, i.e. it has a polynomial
dependence on $\bp$. The polynomial operator $\hat{O}_{\bp} = \frac{\partial}{\partial
\bp}$ has the promised property that $\hat{O}_{\bp}\;\Lambda_\mu(\bp,t) = 0$.
In this example, the commutator actually vanishes
because the delta function forces $x=y$, so that
$\delta^3(\bx-\by)\; \left(\: \bar{u}(x)\gamma_5\gamma_0\gamma_5u(y)-
\bar{u}(y)\gamma_5\gamma_0\gamma_5u(x)\: \right) = 0$;
implying that $\Lambda_\mu=0$, and 
one can take $\hat{O}_{\bp}=1$. There does, however, exist examples
where $\hat{O}_{\bp}$ is not trivial as in this case. 
For example, one can show that replacing $b(x)=\gamma_5$ by 
$b(x)=\gamma_5 F^{a}_{\mu\nu}(x)\; F^{a\;\mu\nu}(x)$ yields a non--trivial
$\hat{O}_{\bp}$.

\section{Symmetrization Selection Rules at the Hadronic Level}

We now investigate the physical consequences of the FSSR, particularly
the amplitude for a $\{1,3,5\ldots\}^{-+}$ hybrid current to create 
$\eta\pi^0$, which can be obtained from the Green's function.
This is done via an alternative route to the former treatment~\cite{pene},
which contains erroneous aspects~\cite{sim}. 

A natural quantity to obtain from the three--point Green's function is the
amplitude for the hybrid current to create a stable two--body state. 
The hybrid current is not expected to interpolate for a stable particle,
so that we shall not be able to extract the T--matrix for a stable
hybrid particle
to decay to the two--body state. To obtain the amplitude, a 
complete set of asymptotic,
i.e. stable, states $n$ are inserted in the Green's function. 
The leads to a general relation (Eq. \ref{asy}) 
between the Green's function and
the amplitudes for a $\{1,3,5\ldots\}^{-+}$ hybrid current to create 
the asymptotic states. 
Under certain conditions this equation can be 
simplified to show that the Green's function is proportional to
the amplitude for a $\{1,3,5\ldots\}^{-+}$ hybrid current to create 
$\eta\pi^0$, Eq. \ref{two}.
Because some explicitly identified contributions to 
$\hat{O}_{\bp}\; G_{\mu}(\bp,t)$ vanish, it follows that the amplitude
does not come from these contributions.
 
Restricting to QCD on its own~\cite{own}, from the definition Eq. \ref{def}
\beqn\plabel{ins}
{G_\mu}(\bp,t)=\int\; d^3x\; d^3y\; e^{i(\bp\cdot\bx-\bp\cdot\by)}
\sum_n\; \langle  0 | {B}(\bx,0)\; {C}(\by,0)  \; e^{-iHt} 
|n\rangle \; \langle n|{A_{\mu}}(z) |0\rangle
\eeqn
where we used time translational invariance of the fields $B(x)$ and 
$C(y)$,
e.g. $B(\bx,t) = e^{iHt} B(\bx,0) e^{-iHt}$, with
$H$ the QCD Hamiltonian.
The product ${B}(\bx,0)\; {C}(\by,0)$ and
${A_{\mu}}(z)$ should be colourless to make the expression non--zero,
which is the case since each current has been assumed to be gauge invariant. 
If the quarks have their physical masses, the lowest asymptotic
states contain the states $\pi$ and $\eta$, the lowest stable states of the
QCD spectrum~\cite{int}. The $\pi$ only decays weakly and
electromagnetically, and is hence stable as far as QCD 
on its own is concerned. The $\eta$ has a width
of only $10^{-5}$ times the typical hadronic width of $100$ MeV, so that
it is very nearly stable.  

Space translational invariance of the fields $B(x)$ and $C(y)$, 
e.g. $B(\bx,0) = e^{-i{\bf P}\cdot\bx} B(0) e^{i{\bf P}\cdot\bx}$, with
${\bf P}$ the QCD momentum operator, is now employed. 
Performing one of the integrations
in Eq. \ref{ins} (the one over $\bx+\by$)

\beqn\plabel{asy1}
{G_\mu}(\bp,t)=  \sum_n\; (2\pi)^3\; \delta^3(\bp_n)\; e^{-iE_nt}\; 
\langle  0 |{(}\int d^3x\; e^{i\bp\cdot\bx}\; {B}(\bx,0){)} \; {C}(0)
|n\rangle \; \langle n| {A_{\mu}}(z)|0\rangle
\eeqn
where the integration variable $\bx-\by$ is denoted by $\bx$.
Here $\bp_n$ and $E_n$ are the momentum and energy of state $n$.
The Euclidean space analogue of all the steps up to here
is obtained by taking $t\rightarrow -it$. However, we now restrict
to Minkowski space in order to integrate over time.
From Eq. \ref{asy1} 

\beqna\plabel{asy}\lefteqn{\hspace{-14cm}\int^{\infty}_{-\infty} dt\; 
\hat{O}_{\bp}\;G_\mu(\bp,t)\;
e^{iEt} \nonumber } \\
=  \sum_n\; (2\pi)^4\; 
\delta^3(\bp_n)\; \delta(E_n-E)\;
\hat{O}_{\bp}\;\langle  0 |{(}\int d^3x\; 
e^{i\bp\cdot\bx}\; {B}(\bx,0){)} \; {C}(0)
|n\rangle \; \langle n| {A_{\mu}}(z)|0\rangle 
\eeqna
where $E$ is a real number.
The delta functions indicate that the asymptotic states are at rest and
have energy $E$.

For $E$ below the $\pi^+\pi^-\pi^0$ or $\pi^0\pi^0\pi^0$
(called the $\pi\pi\pi$) threshold only $\pi\pi$ asymptotic states
contribute. Since for two--pion states at rest
$\langle \pi^0\bk\;\pi^0 -\bk| {A_{\mu}}(z)|0\rangle$
and $\langle \pi^+\bk\;\pi^- -\bk| {A_{\mu}}(z)|0\rangle$
respectively vanish by Bose symmetry and $CP$
conservation, this forces the L.H.S. of Eq. \ref{asy} to be zero.
If $E$ is above the $\pi\pi\pi$ threshold the sum in Eq. \ref{asy}
can be shown to be infinite. However, we shall 
show below that if the $\eta\pi^0$ threshold can be made below the
$\pi\pi\pi$ threshold, in contrast to  
experiment,  the sum has only one contribution. This is 
important because we want a {\it single} amplitude on the R.H.S. of 
Eq. \ref{asy} to be proportional to the L.H.S. 


In QCD we are free to tune the quark masses away from the masses
corresponding to experiment. Taking the $\eta\pi^0$ threshold to be below
the $\pi\pi\pi$  threshold for some range of quark masses, is called the 
``QCD dynamics'' condition.
This condition can heuristically be satisfied by
noting that four times higher up / down quark masses would
yield a two times heavier $\pi$, while the $\eta$ mass would not change 
much, so that $\eta$ can move below the $\pi\pi$ threshold. 
This follows from the fact that the $\pi$ and $\eta$ masses,
$m_\pi$ and $m_\eta$, are
$c\sqrt{m_u+m_d}$ and $c\sqrt{(4m_s+m_u+m_d)/3}$ by chiral 
symmetry breaking, with $m_{u,d,s}$ the current quark masses and 
$c$ a constant. Under the QCD dynamics condition $\eta$ lies below
$\pi\pi$ threshold, so that it is stable. 
This means that the formalism is exact when asymptotic states 
involving $\eta$ are inserted in Eq. \ref{ins}.

Consider $E$ between the $\eta\pi^0$ and $\pi\pi\pi$ thresholds. Then the 
sum in Eq. \ref{asy} is over all on--shell $\pi$ and $\eta$ states with
momenta $\bk_1$ and $\bk_2$ respectively, i.e. $\sum_n = \int d^3 k_1 \;
/(2\pi)^3\; \int d^3 k_2\; /(2\pi)^3$. 
Noting that $\bp_n = \bk_1 + \bk_2$ and $E_n = \sqrt{\bk_1^2+m_\pi^2}
+ \sqrt{\bk_2^2+m_\eta^2}$, and performing the integrations, Eq. \ref{asy}
implies that

\beqna\plabel{two}\lefteqn{\hspace{-0.8cm}\int^{\infty}_{-\infty} dt\; 
\hat{O}_{\bp}\;G_m(\bp,t)\; e^{iEt} = f(q^2)\;  
\frac{1}{8(2\pi)^2E^4}\;\sqrt{(E^2-(m_\pi+m_\eta)^2)(E^2-(m_\pi-m_\eta)^2)}
\; \nonumber } \\ & & \hspace{-1.6cm}\eqntimes 
(E^4-(m_\pi+m_\eta)^2(m_\pi-m_\eta)^2) 
\int d\Omega_{\bk}\;  q_{m} \;
\hat{O}_{\bp}\;\langle  0 |{(}\int d^3x\; 
e^{i\bp\cdot\bx}\; {B}(\bx,0){)} \; {C}(0) 
|\pi\bk\; \eta -\bk\rangle 
\eeqna
Here $q^{\mu} \equiv k_1^{\mu} - k_2^{\mu}$ and $\bk_1 = -\bk_2 \equiv \bk$.
In the expression 
$q^2$ is a function of $|\bk|$, and $|\bk|$ is restricted by $E_n=E$,
so that $q^2$ is a function of $E$.
The amplitude for a hybrid current to create $\eta$ and $\pi$ mesons,
$\langle\pi\bk\; \eta -\bk| {A_{\mu}}(z)|0\rangle$, should occur in Eq. 
\ref{two}. However, we used the Lorentz properties of 
$\langle\pi\bk\; \eta -\bk| {A_{m}}(z)|0\rangle$ to define it as
$q_{m}f(q^2)$, with $m$ a spatial index, 
for $A_{\mu}$ having $J=1$~\cite{ind}. For a time index $m$ and other $J$ 
similar results follow. 


On the R.H.S. of Eq. \ref{two}, $\langle  0 | {B}(\bx,0) \; {C}(0) 
|\pi\bk\; \eta -\bk\rangle$ is a property of $\eta$ and $\pi$.
For example, it contains a term proportional to
$\langle  0 | {B}(0)|\pi\bk\rangle \; 
\langle  0 | {C}(0)|\eta \bk\rangle$, a product of individual
properties of $\pi$ and $\eta$, when the currents
$B(0)$ ($C(0)$) have the same quantum numbers
as the asymptotic states $\pi$ ($\eta$). 
One hence interprets Eq. \ref{two} as meaning that the decay described
by the three--point Green's function on the L.H.S., is proportional 
to the decay amplitude described by $f(q^2)$ on the R.H.S., up to 
a ``constant'' of proportionality which describes properties of 
$\eta$ and $\pi$. Suppression of the decay on the L.H.S. should hence
translate into suppression of the decay on the R.H.S., since the properties
of $\eta$ and $\pi$ should remain unaltered. 

From Eq. \ref{two}, noting that $f(q^2)$ is proportional to the L.H.S.,

{\bf Symmetrization selection rules at the hadronic level (SSR):} {\it 
The amplitude for a neutral $J^{PC}=\{1,3,5\ldots\}^{-+}$ exotic
hybrid meson current
to create or annihilate $\eta\pi^0$, $q_m f(q^2)$,
 does not arise, in QCD, from contributions to $
\hat{O}_{\bp}\;G_m(\bp,t)$ that vanish
by the FSSR. This holds for quark masses chosen such that the
$\eta\pi^0$ threshold is below the $\pi^+\pi^-\pi^0$ or
$\pi^0\pi^0\pi^0$ thresholds, and for $E$ between these thresholds.}

This concludes the physical consequences of the FSSR.

\section{Large $N_c$ \plabel{sec5}}

We now study the effect of taking $N_c$ to 
be large in QCD~\cite{cohen}. 
Here one makes a classification
based on power counting in $N_c$.
The amplitude $\langle\pi\bk\; \eta -\bk| {A_{\mu}}(z)|0\rangle$
is ${\cal O}(1)$~\cite{lebed}, so that $f(q^2)$  is
${\cal O}(1)$.
The contributions to the Green's function from topology 1 is 
${\cal O}(N_c)$ to leading order,
from topologies 2 and 3b ${\cal O}(1)$ and from 
topology 3a ${\cal O}(\frac{1}{N_c})$
\cite{lebed}, so that the L.H.S. of Eq. \ref{two} would ordinarily be
${\cal O}(N_c)$. However, since topology 1 does not contribute
by the FSSR, the L.H.S.
is ${\cal O}(1)$, and hence $f(q^2)$ is ${\cal O}(\frac{1}{N_c})$ 
\cite{ext}. 
Hence both quantities are subleading to their usual $N_c$ counting.
Whence

\beqn
\frac{1}{N_c}\;\hat{O}_{\bp}\; G_\mu(\bp,t) \rightarrow 0 
\hspace{1.5cm} f(q^2) \rightarrow 0 
\hspace{1.5cm}\mbox{as   } N_c  \rightarrow\infty
\eeqn
where both expressions ordinarily remain non--zero as $N_c \rightarrow\infty$.
The two exactly vanishing 
expressions are respectively the 
consequences of the FSSR and SSR in large $N_c$. 
The SSR in large $N_c$ is:
{\it The hybrid $J^{PC}=\{1,3,5\ldots\}^{-+}\rightarrow\eta\pi^0$ 
amplitude is ${\cal O}(\frac{1}{N_c})$ in QCD, i.e. vanishes exactly 
in the large $N_c$ limit},
for quark masses chosen such that the
$\eta\pi^0$ threshold is below the $\pi^+\pi^-\pi^0$ or
$\pi^0\pi^0\pi^0$ thresholds, and for $E$ between these thresholds. 
This is {\it not} the same as Bose symmetry, since $\eta$ and $\pi^0$
are not identical particles in the large $N_c$ limit.
The finding that the amplitude for 
the hybrid current to decay to $\eta\pi^0$ is subleading in $N_c$ 
intuitively follows from the fact that OZI--violating processes are 
large $N_c$ suppressed.


The preceding discussion in this section assumed that the L.H.S. of
Eq. \ref{two} is ``ordinarily'' of ${\cal O}(N_c)$. This is attained
when $\langle  0 | {B}(\bx,0) \; {C}(0) 
|\pi\bk\; \eta -\bk\rangle$ on the R.H.S. of Eq. \ref{two}
is ${\cal O}(N_c)$. This is true when the
currents $B(x)$ ($C(y)$) individually have the same quantum numbers
as the asymptotic states $\pi$ ($\eta$) or $\eta$ ($\pi$)~\cite{lebed},
i.e. when the currents are pseudoscalar. Up to this point in the
Paper we have not
required either of the currents to have a specific parity and charge
conjugation. However, if the currents are not pseudoscalar,
$\langle  0 | {B}(\bx,0) \; {C}(0) 
|\pi\bk\; \eta -\bk\rangle$ is at most
${\cal O}(\sqrt{N_c})$~\cite{lebed}, so that the L.H.S. cannot be
${\cal O}(N_c)$, and hence should be regarded as ordinarily 
${\cal O}(1)$: the next possibility in the $N_c$ counting. 
The FSSR still yield that
the  L.H.S. is ${\cal O}(1)$, but as this is not new information, the
FSSR imply no additional restrictions on $f(q^2)$. Thus the SSR in 
large $N_c$ are only interesting when $B(x)$ and $C(y)$ are 
pseudoscalar currents.

This concludes the statements of the FSSR and SSR in large $N_c$.
A few final remarks are in order.

\section{Remarks}

Firstly, on the relationship between $SU(3)$ and large $N_c$.
For equal mass up, down and strange quarks ($SU(3)$ flavour symmetry) the
$\eta$ and $\pi$ are among the degenerate lightest states of QCD, 
satisfying the QCD dynamics condition that the $\eta\pi^0$ threshold is
below the $\pi\pi\pi$ threshold, 
implying the existence of SSR. 
The SSR and Eq. \ref{two} yield that the hybrid 
$J^{PC}=\{1,3,5\ldots\}^{-+} \rightarrow\eta\pi^{0}$ amplitude vanishes
exactly because topologies 2 and 3a 
in Fig. \ref{topo} vanish in the $SU(3)$ limit, noting that an $SU(3)$ octet
current $B(x)$ or $C(y)$, interpolating respectively for 
$\eta$ or $\pi^0$, does not couple to a quark--antiquark pair created
from the vacuum.
It has been known independently
 for some time that the $SU(3)$ octet $\{1,3,5\ldots\}^{-+}
\rightarrow\eta\pi^{0}$ amplitude vanishes exactly in the $SU(3)$ 
limit~\cite{sel,meshkov}.

Hence the hybrid 
$\{1,3,5\ldots\}^{-+} \rightarrow\eta\pi^{0}$ amplitude
vanishes exactly in {\it either} the large $N_c$ or $SU(3)$ limits, but
the one does {\it not} follow from the other, as $SU(3)$ symmetry 
does not derive from, or does not imply, the large $N_c$ limit~\cite{lebed}. 
The amplitude should be more suppressed than
either limit indicates, due to the constraints from the other limit.  

Secondly, on the role of Bose symmetry.  
One finds that contributions to the Green's function that vanish
by a field theoretical version of Bose symmetry, at least after the
polynomial operator is applied, do not contribute to the hybrid 
$\{1,3,5\ldots\}^{-+}\rightarrow\eta\pi^0$ amplitude. 
However, this amplitude does not itself vanish by Bose symmetry.

We now remark on the field symmetrization selection rules.
The vanishing contributions to the Green's function were
foreshadowed by, and have direct analogues in, 
the  ``symmetrization selection rules I'' of
the non-- field theoretic analysis of ref.~\cite{sel}, where decays of
$\{1,3,5\ldots\}^{-+}$ hybrids
to two $J=0$ (hybrid) meson states which are identical in all respects except 
possibly flavour are prohibited.
The latter condition is translated into the requirement that $b(x) = c(x)$
for the FSSR. In ref.~\cite{sel} the selection rule applied 
to flavour components of the (hybrid) 
mesons B and C which are identical, e.g. 
$\bar{u}u$ for both. This is exactly the case for the contributions to the
Green's function for which we have FSSR.

The symmetrization selection rules at the hadronic level have important
consequences for models. The hybrid 
$\{1,3,5\ldots\}^{-+}\rightarrow \eta\pi^0$ amplitude only comes
from disconnected topologies. This feature puts the SSR in 
contradiction with most current models of QCD, although
most find it in an approximate form. Particularly, the 
flux--tube and constituent gluon models all
find an approximate selection rule for the connected decay of the low--lying
$1^{-+}$ hybrid to $\eta\pi^{0}$~\cite{sel}. In these
models the decay is proportional
to the difference of the sizes of the $\eta$ and $\pi$ wave functions,
which is in contradiction with the SSR. 
Constituent gluon models also have low--lying hybrids 
called ``quark excited'' hybrids whose connected decay
to $\eta\pi^{0}$ vanishes exactly, consistent with the 
SSR~\cite{safir}.
A non--zero decay $1^{-+}\rightarrow\eta\pi^0$ via final state interactions
has been estimated from the decay of $1^{-+}$ to two mesons
which then rescatter via meson exchange to $\eta\pi^0$~\cite{don}.
The process is described in QCD by connected decay
 (with a quark loop), so that
it contradicts the SSR.
In practical calculations in the above models the QCD dynamics condition
may not be satisfied, so that the model calculations
are strictly not required to 
obey the SSR. However, model parameters can be changed to satisfy the
QCD dynamics condition, so that the SSR have to be obeyed. Hence models
that do not incorporate vanishing connected topologies are inadequate.
In QCD sum rules, the connected topology vanishes,
consistent with the FSSR; and $f(q^2)$ is small, 
consistent with expectations from the OZI rule~\cite{qcdsr}. 
The QCD sum rule calculations~\cite{qcdsr} constitute explicit examples
of the results of this work. 

If one's aim is to obtain information about the physically interesting
hybrid $\{1,3,5\ldots\}^{-+} \rightarrow\eta\pi^{0}$ amplitude, it
is possible to do so from a variety of Green's functions. The essential 
point about a quantum field theory approach is that information about
amplitudes can be extracted from various Green's functions. In addition
to Green's functions involving a fermion--antifermion current going to
two fermion--antifermion currents considered in this Paper and elsewhere
\cite{pene,qcdsr}, Green's functions containing a 
quark--antiquark current going to a pure glue and 
a quark--antiquark current have been considered in QCD sum rules~\cite{qcdsr}.
The latter case can be shown to yield no selection rules for
Green's functions (FSSR), and hence cannot be used
to deduce selection rules for physical amplitudes (SSR). 
However, the fact that the SSR cannot be deduced does
not imply that the SSR is not valid.

We developed SSR for the hybrid 
$\{1,3,5\ldots\}^{-+} \rightarrow\eta\pi^{0}$ amplitude.
Related SSR have been found for four--quark and glueball initial 
states~\cite{sel}. These SSR are derived in a formal context which
can transparently be extended to a rigorous quantum field theoretic 
argument along the lines of this work. Extension to two--body final states
beyond $\eta\pi^0$ is more conceptually involved, due to the analogue
of the QCD dynamics condition. In the interest of brevity these extensions
should be considered in future work.

Lastly, we remark on experimental consequences.
The Crystal Barrel experiment has recently claimed evidence for
$p\bar{p}\rightarrow \mbox{ resonant } 1^{-+}\rightarrow \eta\pi^0$ at a 
level that
is a significant fraction of especially the P--wave $p\bar{p}$ 
annihilation~\cite{cbar}. Although the branching fraction of 
$1^{-+}\rightarrow \eta\pi^0$ is not known, it is reasonable to assume that
it is substantial. This is qualitatively 
at odds with the SSR if the $1^{-+}$ resonance is interpreted
as a hybrid. 
Even though this is not rigorous, as the QCD dynamics
condition is not satisfied experimentally, the small change in 
quark masses from their experimental values needed to enable the validity of 
the QCD dynamics condition indicates
that the $1^{-+}$ resonance
is qualitatively inconsistent with being a hybrid meson.
Other possibilities for the interpretation of the 
$1^{-+}$ enhancement have recently been discussed in refs. 
\cite{sel,safir,don,ach}.



\section*{Acknowledgements}

Useful discussions with A. Blotz, T. Cohen, T. Goldman, E. Golowich, 
R. Lebed, A. Leviatan, 
L. Kisslinger, K. Maltman, M. Mattis, M. Nozar and G. West are acknowledged.
This research is supported by the Department of Energy under contract
W-7405-ENG-36.

\section*{Appendix}

We show that $\Lambda_\mu$ defined in Eq. \ref{uit} has a polynomial 
dependence on $\bp$ if the number of derivatives acting on individual
fields in the currents $B(x)$ and $C(y)$ is bound. 

The commutator in Eq. \ref{uit}
can be expressed, by a general property of commutators,
as a sum of terms, each of which
contains either a commutator of two boson (gluon or photon) fields, or an
anticommutator of fermion $\psi$ 
and conjugated fermion $\bar{\psi}$ (quark or lepton) fields.
Here we used the fact that boson fields commute with fermion fields, that two
fermion fields anticommute, and that two conjugated fermion fields
anticommute. 
Each of the
(anti)commutators is proportional to delta functions 
$\delta^3(\bx-\by)$ (or derivatives acting
on delta functions), by virtue of the canonical (anti)commutation relations
of fields at equal time $t$. 
For example, for commutators of photon fields
$[A_\mu(\bx,t),A_\nu(\by,t)]=0$, 
$[\dot{A}_\mu(\bx,t),A_\nu(\by,t)]=ig_{\mu\nu}\delta^3(\bx-\by), \ldots$;
and for anticommutators of lepton and conjugated lepton fields
$\{\psi_\xi(\bx,t),\bar{\psi}_\sigma(\by,t)\}=
\gamma^0_{\xi\sigma}\delta^3(\bx-\by)$,
$\{\dot{\psi}_\xi(\bx,t),\bar{\psi}_\sigma(\by,t)\}=
-\vec{\gamma}\cdot\vec{\partial}_{\bx}\delta^3(\bx-\by), \ldots$.
Spacial derivatives $\vec{\partial}_{\bx}$
and $\vec{\partial}_{\by}$ might be acting on these (anti)commutators,
since $b(x)$ in Eq. \ref{uit} will in general
 contain derivatives acting on fields.
A derivative $\vec{\partial}_{\by}$ can be expressed in terms of a
derivative $\vec{\partial}_{\bx}$ since the delta function only depends
on $\bx-\by$.
The possibility of temporal derivatives acting on the fields
have already been incorporated
in the (anti)commutation relations. 
Hence a generic term contributing to Eq. \ref{uit} is of the form

\beqn\plabel{bound}
\int\; d^3x\; d^3y\; e^{i(\bp\cdot\bx-\bp\cdot\by)} 
\; f_\mu(x,y,z) \; \vec{\partial}_{\bx}^n \delta^3(\bx-\by) 
\eeqn

One now performs integration by parts over the variable $\bx$, 
which yields powers of
$\bp$ when the derivatives act on the exponential, as well as
derivatives acting on $f_\mu(x,y,z)$. 
(There is no surface term as the delta function does not contribute
for $\bx$ far from $\by$). Eventually there will be no
derivatives acting on the delta function. When performing one of the
integrations, the delta function forces $e^{i(\bp\cdot\bx-\bp\cdot\by)}=1$,
so that the only $\bp$ dependence is the various powers of $\bp$.
Hence $\Lambda_\mu$ has a polynomial dependence on $\bp$. 
This is true as long as the number of derivatives
$n$ in Eq. \ref{bound} is bound.

\end{document}